\documentclass[twocolumn,modern]{openjournal}
\usepackage{epsfig}
\usepackage{amsmath}
\usepackage{hyperref}
\hypersetup{
 colorlinks,
citecolor = [rgb]{0,0.5,0} }
\usepackage{amsmath}
\def\gtwid{\mathrel{\raise.3ex\hbox{$>$\kern-.75em\lower1ex\hbox{$\sim$}}}}
\def\ltwid{\mathrel{\raise.3ex\hbox{$<$\kern-.75em\lower1ex\hbox{$\sim$}}}}
\def\\{\hfil\break}

\def\eg{{\it e.g.}}

\def\sun{\hbox{$\odot$}}
\def\lesssim{\mathrel{\hbox{\rlap{\hbox{\lower2pt\hbox{$\sim$}}}\raise2pt\hbox{$<$}}}}
\def\gtrsim{\mathrel{\hbox{\rlap{\hbox{\lower2pt\hbox{$\sim$}}}\raise2pt\hbox{$>$}}}}

\def\arcdeg{\hbox{$^\circ$}}

\newcommand{\unit}[1]{\ifmmode \:\mbox{\rm #1}\else \mbox{#1}\fi}
\begin{document}

\title{A low value of $H_0$ in tension with the distance ladder from Megamasers using peculiar velocity reconstruction}

\author{Richard Watkins}
\email{rwatkins@willamette.edu}
\affiliation{Department of Physics, Willamette University, Salem, OR 97301, USA}

\author{Hume A. Feldman}
\email{feldman@ku.edu}
\affiliation{Department of Physics \& Astronomy, University of Kansas, Lawrence, KS 66045, USA}

\begin{abstract}
Megamaser distances can be measured without using the traditional distance ladder.  As such, they provide an independent probe of the local expansion rate, quantified by the Hubble constant $H_0$.  We re-analyze megamaser distance and redshift data to obtain a new, more accurate, estimate of $H_0$.  The main improvement in our analysis is the use of a new velocity field reconstruction to correct redshifts for peculiar velocities.  This reconstruction utilizes constrained cosmological simulations in addition to redshift survey data to more accurately capture both coherent flows as well as nonGaussian individual galaxy motions.   We argue that previous estimates of $H_0$ from megamaser data were susceptible to bias due to peculiar velocities.    By correcting for peculiar motions using this new reconstruction we obtain a value of $H_0$ that is more consistent with the value from the Cosmic Microwave Background (CMB) than that of the distance ladder, although consistent with both, whereas previous results showed consistency with the distance ladder and tension with the CMB at greater than $2\sigma$.  Our results suggest that the Hubble tension may be due to some heretofore undetermined systematic in the distance ladder rather than new physics.  
\end{abstract}

\
\section{Introduction}
\label{sec:intro}

The Hubble constant, $H_0$, a measure of the current rate of expansion of the Universe, is one of the most fundamental parameters of the standard cosmological model.   Measurements of $H_0$ with the smallest uncertainty are from the cosmic microwave background (CMB), $H_0= 67.4\pm 0.5$ km/s/Mpc \citep{PlanckCosPar16}, and from the local distance ladder, $H_0= 73.5\pm 0.81$ km/s/Mpc \citep{CanCruCso26}.   The disagreement between these two measurements is known as the ``Hubble tension", and is currently considered to be one of the most pressing problems in cosmology.  Many ideas have been proposed to explain the Hubble tension, but it is currently considered to be unresolved (for a review see \citet{CosmoVerse25}).    While the Hubble tension is often framed in terms of early Universe measurements vs. late Universe measurements \citep{VerTreRie19}, some researchers have instead looked at the tension in terms of distance ladder vs. non-distance ladder measurements \citep{PanPer26,Peri24}.   Indeed, the distance redshift relation appears to be consistent with the standard cosmological model from the epoch of recombination ($z\approx 1090$) through to the local Universe ($z\sim 0$) \citep{WatTruFel26}.   This impressive agreement suggests that the Hubble tension cannot be reconciled by invoking new physics in order to change the distance redshift relation, but instead may be connected to how the distance ladder is calibrated \citep[see \eg][]{CueVerRie15}.  

One crucial piece of evidence in determining if this framing is correct is the distances and redshifts of megamasers \citep[][hereafter P20]{PesBraRei20}.  Since megamaser distances can be measured accurately and directly, independent of the distance ladder, they provide a powerful test of whether the Hubble tension can be ascribed to a systematic in the distance ladder or whether the large value of the Hubble constant is indeed a characteristic of the local Universe.  In finding a value of $H_0$ from megamasers alone that agrees with that from the distance ladder, P20 shifts attention away from systematics and towards solutions of the Hubble tension that employ new physics in the early Universe.  

In this paper we challenge the results of P20.  In particular, we argue that the modeling of peculiar velocities as Gaussian noise makes their results susceptible to bias from both coherent motions and nonGaussian individual galaxy motions.   Coherent motions induce correlations between peculiar velocities that can lead to a systematic bias.  NonGaussian tails in the distribution of galaxy velocities can lead to individual peculiar velocities that are much larger than expected in a Gaussian model.  In a small sample, such as the six megamaser galaxies, a large individual peculiar velocity can bias the value of $H_0$ all by itself.    We demonstrate that when an accurate, high resolution reconstruction of the velocity field \citep[][hereafter M25]{McaJasAta25} is used to correct the redshifts of the megamaser galaxies, we obtain a Hubble constant value that is closer to the CMB value than that of the distance ladder.  Our results suggest that the Hubble tension may indeed be due to a systematic hidden within the distance ladder rather than an indication of new physics.  

The ``standard sirens" of gravitational waves from black hole mergers also provide a way to measure the Hubble constant in the local Universe independent of the distance ladder and similarly require peculiar velocity corrections \citep{BlaTur25,MukLavBou21}.  It should be noted that standard siren estimates of $H_0$ also give values closer to that the CMB result, although the uncertainties on these estimates are somewhat larger than for megamasers, and hence are not conclusive.

In Section~\ref{sec:dvp} we discuss the peculiar velocity corrections to the megamaser redshifts. The distance--redshift curve fitting is discussed in Section~\ref{sec:fit}. Section~\ref{sec:res}  gives our results and we conclude with Section~\ref{sec:dis}.

\section{Peculiar Velocity Corrections to Megamaser Redshifts}
\label{sec:dvp}

The P20 megamaser sample consists of six galaxies all at redshifts $z <0.04$.  As such, peculiar velocities make a significant contribution to their measured redshifts and must be accounted for to obtain an accurate measurement of the Hubble constant.  In analyzing this sample to obtain an estimate of $H_0$, P20 primarily modeled peculiar velocities as being random Gaussian distributed noise \citep[see also][]{BarRamDes25}.  This model has the serious drawbacks of 1) not accounting for coherent motions, and 2) of not capturing the nonGaussian tails of the peculiar velocity distribution due to infall into nonlinear structures \citep{WatFel15a,MaTaySco2013,SheDia2001,Sheth96,RaySomSas96,BahGraCen94}.  The Gaussian model does not attempt to correct for the bias caused by peculiar velocities, but rather accounts for peculiar velocities as an additional form of noise; even then it doesn't properly account for the prevalence of peculiar velocity outliers that can have a significant effect on the outcome of the analysis \citep{UpaSaiTar26,BlaTur24,TurBla23}.  A big improvement over the Gaussian velocity model is to use a reconstruction of the local peculiar velocity field.  A velocity reconstruction can predict a peculiar velocity for each galaxy, thus correcting for the bias in the estimate of $H_0$ while also reducing the uncertainty.  Indeed, P20 also investigated using a velocity reconstruction \citep[hereafter C15]{CarTurLav15} to correct for peculiar velocities (method (5) in their Table 2).  However, in using the velocity reconstruction of C15, they located the megamasers in the reconstruction box using the measured redshift only, without accounting for the peculiar velocity as intended by C15.  This reconstruction uses the galaxy distribution from the 2M++ redshift survey together with gravity to predict peculiar velocities.  It is notable that this method yielded a value of $H_0$ which was on the low end of the methods they looked at, although it differed from their main result by less than $1\sigma$.  As we shall show below, when C15 velocities are used as intended, the resulting $H_0$ is more consistent with the CMB than with the distance ladder.

While it is typical to think of peculiar velocity corrections as accounting mostly for coherent motions that can bias determinations of $H_0$,  in analyzing the small maser sample, especially with its wide range of uncertainties, we need to consider that each individual peculiar velocity correction can have a big impact on the result, whether it is part of a coherent flow or not.  This is due to the effect of random motions not averaging out as they would in a larger sample.  The susceptibility of the analysis to individual velocities, together with the nonGaussian distribution of peculiar velocities, suggests that both coherent and random motions could significantly bias a measurement of $H_0$ such as that made by P20.  In order to obtain an accurate estimate of $H_0$ from the megamaser sample, then, we should use the most accurate velocity reconstruction available, preferably one that has the resolution to capture nonlinear motions \citep[for an alternative to this method that doesn't assume a cosmological model see][]{UpaSaiTar26}.   

\begin{table*}[ht]
  \centering
   \caption{Parameters for the megamaser galaxies taken from P20, including peculiar velocities from C15 using both the method of P20 and the method of C15. We also show the averages and standard deviations of the peculiar velocities from the 80 M25 velocity field realizations.}
 \begin{tabular}{cccc | cccc}
    \hline 
    &&&&\multicolumn{4}{c}{Peculiar Velocity} \\
    \hline
    Galaxy & d (Mpc)& $\sigma_d$ (Mpc) & $cz$ (km/s) & C15-P20 (km/s) & C15(km/s) &Avg. M25 (km/s) & Std. Dev. M25 (km/s)\\
    \hline
   NGC 4258& 7.58& 0.11 &679.3 & 262 & 267 &301 & 99\\
    UGC 3789 & 51.5& 4.2& 3319.9  & -54 & -43 & 76& 54\\
    CGCG 074-064 & 87.6&7.5& 7172.2 &297 & 432& 395& 128\\
    NGC 6323  & 109.4& 30&7801.5 & 414 & 397 & 650 & 54\\
    NGC 5765b  & 112.2& 5& 8525.7 & 124 & 367 & 260&203 \\
    NGC 6264 & 132.1& 20& 10192.6 & 224 & 294 & 293& 196\\
    \hline
  \end{tabular}
 \label{ta:par}
\end{table*}

Since the analysis of P20, a new velocity field reconstruction has been developed by M25, which uses the 2M++ together with cosmological simulations.   This reconstruction is a significant improvement over C15 \citep{StiDesDev26}.   In particular, with the addition of constrained cosmological simulations, M25 is able to resolve the velocity field more accurately and on somewhat smaller scales than C15, crucially including small--scale nonlinear galaxy motions  \citep{WatTruFel26,WatClaFel26,McaJasAta25}.   Using this velocity reconstruction to account for the peculiar velocities of the megamaser galaxies should yield a more accurate estimation of $H_0$ than those obtained by previous methods.  

The main challenge with using velocity reconstructions is locating the objects of interest within the reconstruction grid.  Given that velocity reconstructions derive the density field from a redshift survey, it seems reasonable to locate objects using their observed redshifts.  Indeed, since the reconstruction gives peculiar velocities at each position, it is possible to calculate the observed redshift at each grid position to more accurately place object by their observed redshift.  This is the method used in C15.  A problem that can arise in using this method is that there can be multiple positions in the grid that correspond to the same observed redshift, particularly around clusters.  However, this problem does not occur in the C15 reconstruction for any of the megamasers.   

In using C15 to find the peculiar velocities of the megamasers, P20 chose to locate objects in the reconstruction as if the observed redshift was the cosmological redshift, thus using a different method than advocated by C15.  In Table~\ref{ta:par}, we show the parameters of the 6 megamaser galaxies drawn from P20; for distances reported with asymmetric uncertainties,  we have averaged the two sides.  We also include peculiar velocities  found by P20 using the C15 reconstruction, as well as the peculiar velocities found using the method of C15; we see that the resulting peculiar velocities are quite different, particularly for NGC 5765b.  This megamaser is particularly important in the fit for $H_0$ due to its relatively small uncertainty in distance.  

M25 use a different method for locating objects in their reconstruction.  They use observed distance instead of observed redshift, thus avoiding the multiple solution issue described above.  However, using observed distance to locate objects in the velocity reconstruction has several disadvantages as compared to using observed redshift.  First, distance errors can be quite large, particularly for distant objects, introducing substantial additional uncertainty into the reconstructed peculiar velocity.  Indeed, this additional uncertainty can be large enough to greatly reduce the usefulness of the reconstruction.  Second, converting distances in Mpc to the units $h^{-1}$Mpc used in the reconstruction depends on the value of the Hubble constant $H_0$.  While one can use the value of $H_0$ under consideration (say, in a likelihood analysis), having different peculiar velocities for different values of $H_0$ can also introduce additional uncertainty.  

The M25 reconstruction is given as a set of 80 realizations of the velocity field that are plausible representations of the local Universe. In this paper we choose to locate megamasers in the M25 reconstructions using their observed redshifts.  Some of the megamasers have multiple solutions in some of the realizations; here we choose to average over the multiple solutions.   In Table~\ref{ta:par} we give the averages and standard deviations for the megamaser peculiar velocities over the 80 realizations.   
We note that there is a range of standard deviations over the different realizations.  In particular, the largest standard deviations occur for megamasers with multiple solutions for the observed redshift in many of the realizations.   The standard deviation is a good indication of the how well the realization has estimated the peculiar velocity of the megamaser.  Thus for our fiducial model we will take $\sigma_v$ to be different for each megamaser and have a value given by the standard deviation listed in the table.  

\begin{figure}[ht]
\centering
\includegraphics[scale=0.3]{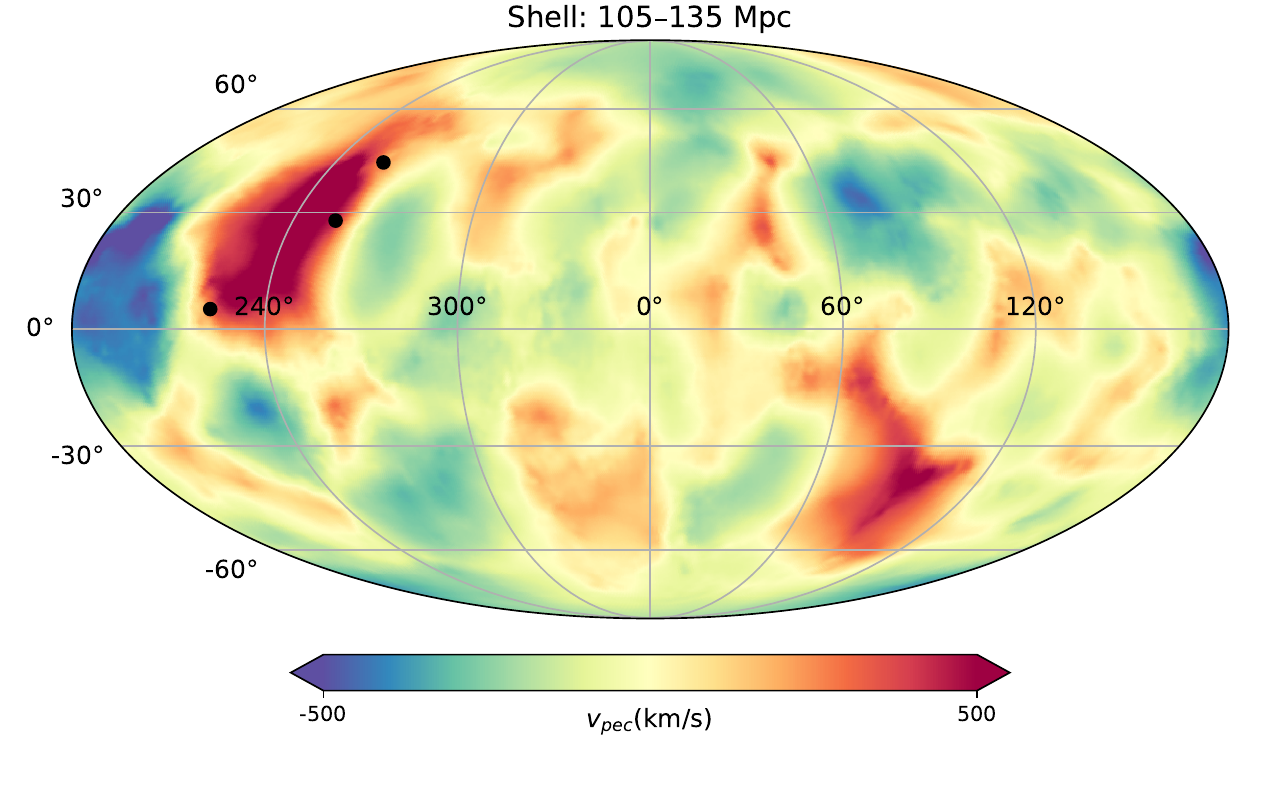}
\caption{The Mollweide projection of one of the M25  velocity field realizations for a shell between 105 and 135 Mpc.  The angular positions of the furthest three megamasers, which are at distances within this shell, are shown as black dots.  We see that the three megamasers participate in a large coherent outflow.  Other M25 realizations give similar results.}
\label{fig:proj}
\end{figure}

Comparing the C15 and M25 velocities, we first note that they are roughly consistent within the standard deviations given, except for NGC 6323.  For this galaxy M25 gives a consistently large value; this is probably due to the finer resolution of M25 in capturing small-scale motion.  
The fact that by chance all of the maser galaxies have positive peculiar velocities means that when these velocities are used to correct the redshifts, they will all become smaller, thus systematically decreasing the estimate of the Hubble constant.  In the case of the maser sample, then,  peculiar velocities have the effect of biasing $H_0$ estimates to be larger than they should be, creating an appearance of consistency with the distance ladder.  

The peculiar velocities of the farthest three galaxies are at least partly due to coherent motions.  In Fig.~\ref{fig:proj} we show the projection of one of the M25 radial velocity field realizations for a shell between 105 and 135 Mpc. The other realizations are similar.   In the figure, we include the angular positions of the furthest three megamasers, which are at distances within this shell, as black dots.  We see that all three of these objects participate in the same large, coherent, outflow.    This suggests that as more megamasers are found in a broader region of the sky the magnitude of the peculiar velocity bias should be somewhat reduced.  

\section{Curve Fitting}
\label{sec:fit}

Our goal is to fit a distance-redshift relation to data by minimizing the likelihood
\begin{equation}
 {\cal L}(H_0)= \prod_i \frac{1}{\sqrt{2\pi}\sigma_i}\exp\left(- \frac{ (d_i - d_{m}(z_i))^2}{2\sigma_i^2}\right)
\label{eqn:chi2}
\end{equation}
where $d_m(z)$ is a model that depends on the value of $H_0$ and potentially other parameters and $\sigma_i$ is the uncertainty for the $i$th object.  For simplicity we will assume flat $\Lambda$CDM, so that angular diameter distance is given by 
\begin{equation}
d_m(z) = \frac{c}{H_0(1+z)}\int_0^z \ \frac{dz^\prime}{\sqrt{\Omega_m (1+z^\prime)^3 + (1-\Omega_m)}}.
\label{eqn:dz}
\end{equation}
Here we will fix the value of the matter density $\Omega_m=0.3$.  For the distance range of the sample, the results are insensitive to the precise value of $\Omega_m$.  Thus our model $d_m(z)$ has essentially only one free parameter, the Hubble constant $H_0$.  

The uncertainty $\sigma_i$ includes the measurement uncertainty in the $i$th distance measurement $\sigma_{di}$, but should also account for the error in the peculiar velocity model.  This introduces an uncertainty in the measured redshift $z_i$ and thus model distance $d_m(z_i)$.  Here we will assume uncertainty in the assigned peculiar velocities to be $\sigma_v$, due to both errors in the velocity field model as well as deviations of individual galaxy velocities from the smoothed velocity field.  This introduces a deviation of the measured redshift of $\sigma_v/c$, and an uncertainty in $d_m(z)$ of approximately $\sigma_v\left(\frac{d\left(d_m(z)\right)}{c\ dz}\right)$.  All together, then, we add the uncertainty in $d_i$ and the uncertainty in $d_m(z_i)$ in quadrature
\begin{equation}
\sigma_i^2= \sigma_{di}^2 + \left(\frac{\sigma_v}{c}\left.  \frac{d(d_m(z))}{\ dz}\right|_{z_i}\right)^2.
\label{eqn:sigi}
\end{equation}
Our approach differs from P20 in that they consider separate likelihoods for the velocities and the distances. They then get the total likelihood by multiplying these together.  As we shall see in Sec.~\ref{sec:res}, our methods give very similar results.  

Given the range of distance uncertainties in the sample, the velocity uncertainty $\sigma_v$ plays an important role in this analysis beyond that of increasing the overall uncertainty in the $H_0$ estimate.  By adding a uniform uncertainty to all of the galaxies, as in Eq.~\ref{eqn:sigi}, less weight is placed on the objects with small $\sigma_{di}$, such as NGC 4258, that would otherwise dominate the analysis.  Thus, as we shall see, the estimate of $H_0$ obtained is somewhat sensitive to the value of $\sigma_v$ used in the analysis.  This dependence of the estimate of $H_0$ on the value of $\sigma_v$ was not explored by P20.  

Given that the M25 velocity constructions are more accurate and have a higher resolution than that of C15, we would expect them to require a smaller value of $\sigma_v$.   The variation among the M25 realizations shown as the rightmost column in Table~\ref{ta:par} suggests that while $\sigma_v$ can be taken as smaller than $150$km/s for some megamasers, for others the peculiar velocity is less well determined.  This suggests using different $\sigma_v$ for each object.  We take as our fiducial model, then, the M25 velocities together with a $\sigma_v$ given by the standard deviations in the rightmost column of Table~\ref{ta:par}.  This model accounts for variations in how well the M25 reconstruction determines the peculiar velocities.  

\begin{figure}[ht]
\centering
\includegraphics[scale=0.5]{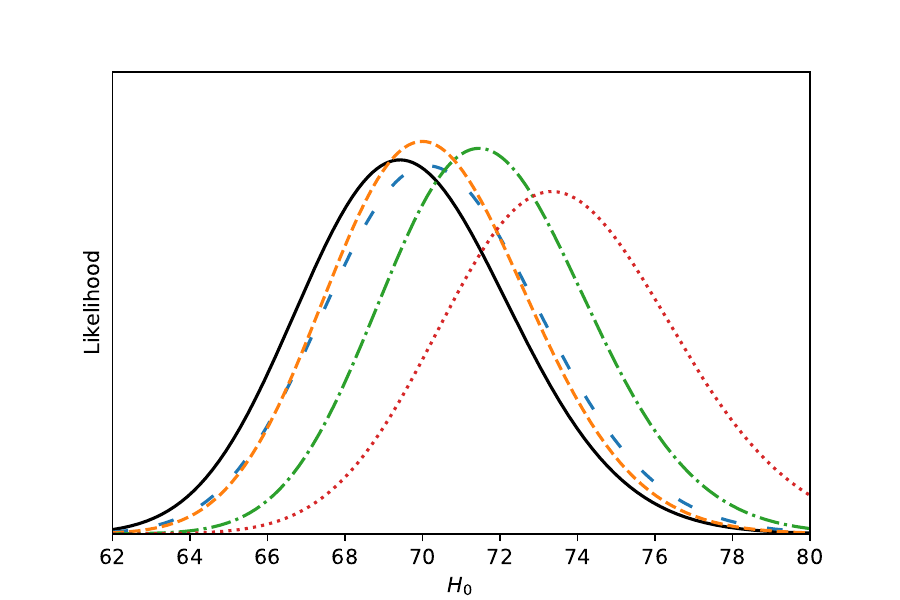}
\caption{Likelihood of $H_0$ for the megamaser data corrected using peculiar velocities directly from C15 (orange dashed curve) and the C15 velocities used by P20 (green dash-dotted curve).  The blue  curve (sparse dashes) is the average of the 80 M25 realizations using a velocity uncertainty of $\sigma_v=150$ km/s.  The leftmost black solid curve is out fiducial model, the average of the 80 M25 realizations using the standard deviations in the last column from Table~\ref{ta:par} as the $\sigma_v$ for each object.  
For comparison, the rightmost red dotted curve is the $H_0$ likelihood for no peculiar velocity correction but with $\sigma_v=250$ km/s, the primary model used in P20. The likelihood curves are normalized to have equal area.}
\label{fig:like1}
\end{figure}

\section{Results}
\label{sec:res}

\begin{table*}[ht]
  \centering
  \caption{Estimates of $H_0$ using various peculiar velocity corrections.  $P(\le H_{CMB})$ is the likelihood that the value of $H_0$ is as small or smaller than that from the CMB, $H_{CMB}= 67.4$ km/s/Mpc, and $P(\ge H_{DL})$ is the likelihood that the value of $H_0$ is as large or larger than that of the distance ladder, $H_{DL}= 73.5$ km/s/Mpc.}
  \begin{tabular}{cccc}
    \hline
    Correction Type & $H_0$ (km/s/Mpc)  & $P(\le H_{CMB})$ & $P(\ge H_{DL})$\\
    \hline
  None $(\sigma_v=250)$ & $73.6\pm 3.0$  & 0.015 & 0.503\\
  C15(P20) $(\sigma_v=150)$ & $71.7\pm 2.7$  & 0.047 & 0.246\\
  C15 $(\sigma_v=150)$& $70.2\pm 2.6$  & 0.137 & 0.110\\
  M25 $(\sigma_v=150)$& $70.4\pm 2.8$ & 0.141 & 0.134\\
  M25 (fiducial) & $69.7\pm 2.7$ & 0.207 & 0.086 \\
    \hline
  \end{tabular}
 \label{ta:vel}
\end{table*}

P20 assume $\sigma_v=150$ km/s for the C15 velocity reconstruction method of correcting for peculiar velocities.   In Fig.~\ref{fig:like1} we show the likelihoods of $H_0$ for the data using the C15 as used by P20 (green curve) and as intended by C15 (orange curve), both using $\sigma_v=150$ km/s. For comparison, we also show the $H_0$ likelihood (red curve) for no peculiar velocity correction but with $\sigma_v=250$ km/s, the primary model used in P20. The estimate of $H_0=71.5\pm 2.6$ km/s/Mpc for the C15 velocities used by P20 matches well with the value given in Table 2 in P20.  This provides a good check that our method gives similar results to those obtained in P20 for this method.   For M25, we have 80 different sets of peculiar velocities for the 6 galaxies.  While it is tempting to calculate the averages of the peculiar velocities given by the realizations for each galaxy, as shown in Table~\ref{ta:par}, this approach would ignore the fact that the six individual galaxy peculiar velocities drawn from a particular realization are strongly correlated.  Instead, we calculate the likelihood for $H_0$ for each realization, and then since each of the realizations is equally likely to be true, we average these likelihoods over the 80 realizations.    Note that this approach accounts for the variations in the realizations in that when we average likelihoods that are peaked in different places we obtain a broader likelihood distribution.   In practice, in spite of the variation in the individual peculiar velocities, the $H_0$ likelihoods for the different realizations are quite similar.  The average of the M25 likelihoods using $\sigma_v=250$ km/s is shown by the blue curve in Fig.~\ref{fig:like1}.  Finally, we show the likelihood for our fiducial model (black curve), the M25 velocities taking $\sigma_v$ to be different for each megamaser and given by the standard deviation of the 80 realizations.

In Table~\ref{ta:vel} we give the maximum likelihood value of $H_0$ (with estimated uncertainty) for the various choices of peculiar velocity corrections.  We also give the probability $P(\le H_{CMB})$, that the value of $H_0$ is as small or smaller than that from the CMB value, $H_{CMB}= 67.4$ km/s/Mpc, and  $P(\ge H_{DL})$, the probability that the value of $H_0$ is as large or larger than that of the distance ladder value, $H_{DL}= 73.5$ km/s/Mpc, both found from the calculated likelihood.  From both Fig.~\ref{fig:like1} and Table~\ref{ta:vel} we see that while treating peculiar velocities as noise gives results consistent with the distance ladder and in some tension with the CMB, when peculiar velocities are corrected for using velocity reconstructions the results tend to be between the two values and consistent with both.  Our fiducial model gives results more consistent with the CMB, but only in light tension with the distance ladder.  

%


\section{Discussion}
\label{sec:dis}

One of the key questions that must be answered in order to understand the origin of the Hubble tension is whether the large expansion rate measured using the local distance ladder is a characteristic of the local Universe or due to some systematic associated with the distance ladder itself.  Megamasers provide a possible way of potentially answering this question, since they are local objects whose distance can be accurately measured independent of the distance ladder.  P20 have measured the Hubble constant using megamasers alone, and by obtaining a value consistent with results from the distance ladder and inconsistent with the CMB value, seemed to answer this question.  However, here we argue that peculiar velocities can bias megamaser determinations of $H_0$ through their impact on the measured redshift, and that  by using an accurate, high resolution,  velocity reconstruction can we avoid these biases.  Our main result is that by using the recent velocity reconstruction of M25, we have obtained a Hubble constant of $69.7\pm 2.7$ km/s/Mpc, closer to the CMB value although consistent with both the CMB and distance ladder values.  Our results are frustratingly ambiguous; clearly it is important to find and measure more megamasers to clarify the situation.

It is notable that the P20 results, treating peculiar velocities as Gaussian random noise, results in an $H_0$ close to the distance ladder value. Correcting redshifts with the peculiar velocities from C15 results in a lower $H_0$, and with M25 reconstructions yields an even lower $H_0$, which is close to the value given by the CMB. Thus our result differs from P20 due to a systematic error caused by peculiar velocities. All six of the megamaser galaxy hosts happen to have a positive peculiar velocity in the M25 velocity reconstructions.  A major contributor to the peculiar velocity bias is that the farthest three megamasers happen to all participate in the same large, coherent outflow.  Both coherent flows and nonGaussian motions are not modeled well by the Gaussian distribution of peculiar velocities assumed by P20.  Together, the positive peculiar velocities of the megamaser galaxies biased the measured redshifts to be systematically low, leading to a Hubble constant estimate that was biased high.  This is already seen in P20, where the method using the C15 velocity reconstruction gave a somewhat smaller estimate of $H_0$ than the other methods.  Using the more accurate and higher resolution M25 velocity reconstruction is even more effective at eliminating the bias, and results in an even smaller estimate of $H_0$ that is more consistent with the CMB than with distance ladder estimates.  

While our work emphasizes the importance of using velocity reconstructions to correct nearby redshifts for peculiar velocities, it also demonstrates how much the particular peculiar velocities obtained from velocity reconstructions depends on the precise way that objects are located in the reconstruction box.   For example, Table~\ref{ta:par} shows how different the megamaser velocities from C15 can depend on whether one locates objects using observed redshift as cosmological redshift or calculates observed redshift for locations in the box using the reconstruction velocities.   Results can be similarly different when objects are located using distance measurements instead of redshift.  As velocity reconstructions improve, more research needs to be done to determine the velocity uncertainty introduced for each method of locating objects in the reconstruction box.  

Our work has several important ramifications.  First, it emphasizes the importance of megamasers as a check on the traditional distance ladder.  Much more work needs to be done in finding galaxies that host megamasers and in improving distance estimation methods before we can definitively determine the value of $H_0$ from this probe of the local Universe.  Second, it highlights peculiar velocities as being a source of bias in low-redshift determinations of the distance redshift relation.  Local velocity reconstructions are an important tool in order to address this bias and should continue to be improved in accuracy and resolution.  Finally, our results suggest that we need to continue to scrutinize the distance ladder and possible systematics in its various rungs.

While we were revising this manuscript we became aware of \citet{StiDes26} who obtain similar results to those reported here.  

\\
\\

\bibliographystyle{mnras}
\bibliography{haf}

\end{document}